# Substrate Integrated Bragg Waveguide: an Octave-bandwidth Single-mode Functional Transmission-Line for Millimeter-Wave and Terahertz Applications

Binbin Hong,[1,2] Naixing Feng,[1] Jing Chen,[1] Guo Ping Wang,[1,*] Viktor Doychinov,[2] Roland Clarke,[2] Nutapong Somjit,[2] John Cunningham,[2] and Ian Robertson[2,†]

[1] *Institute of Microscale Optoelectronics, Shenzhen University, Shenzhen 518060, China*
[2] *School of Electronic and Electrical Engineering, University of Leeds, Leeds LS2 9JT, UK*

We demonstrate an air-core single-mode hollow waveguide that uses Bragg reflector structures in place of the vertical metal walls of the standard rectangular waveguide or via holes of the so-called substrate integrated waveguide. The high-order modes in the waveguide are substantially suppressed by a modal-filtering effect, making the waveguide operate in the fundamental mode over more than one octave. Numerical simulations show that the propagation loss of the proposed waveguide can be lower than that of classic hollow metallic rectangular waveguides at terahertz frequencies, benefiting from a significant reduction in Ohmic loss. To facilitate fabrication and characterization, a proof-of-concept 20 to 45 GHz waveguide is demonstrated, which verifies the properties and advantages of the proposed waveguide. A zero group-velocity dispersion point is observed at near the middle of the operating band. This work offers a step towards a novel hybrid transmission-line medium that can be used in a variety of functional components for broadband millimeter-wave and terahertz applications.

## I. INTRODUCTION

Recently, the millimeter-wave (mmW) and terahertz (THz) frequency bands have attracted much attention owing to their unique applications, such as ultra-broadband spectroscopic sensing [1], high-data-rate wireless communication [2], imaging and non-destructive testing [3], security screening [4] and radio astronomy [5]. However, the development of practical mmW to THz systems has proved to be challenging due to the lack of high-power radiation sources and high-sensitivity detectors, along with the high loss of planar transmission-line structures. Traditional hollow metallic rectangular waveguides are surprisingly difficult to replace as the main electromagnetic waveguiding medium for the integration of subsystems. There is great interest in finding an alternative transmission-line medium that is low-loss, single-mode, broadband, compact, and has good isolation but is much easier to fabricate and suitable for functional components and multichip modules.

Planar transmission lines are widely used in functional components and multichip modules for realizing passive components and interconnects, but they generally suffer from high substrate and radiation losses at 1 THz and above, as well as mode competition with the substrate modes [6]. Reported propagation loss results for microstrip, coplanar waveguide, coplanar stripline, and Goubau line at 1 THz are approximately $\alpha_\mathrm{ms}$=43.3 dB/cm, $\alpha_\mathrm{cpw}$=65.1 dB/cm, $\alpha_\mathrm{cs}$=26 dB/cm, and $\alpha_\mathrm{gl}$ =30.4 dB/cm, respectively [7-10]. The dielectric substrate of planar transmission lines supports unwanted TM modes which limit the single-mode operating frequency band. An on-chip THz liquid sensor based on a Goubau line has been reported operating up to 0.8 THz, the highest operating frequency limit of which is mainly determined by the mode competition between the guided mode and surrounding modes in the substrates or superstrates [11]. The propagation loss of rectangular dielectric waveguide has been reported to be less than 0.54 dB/cm between 0.44 and 0.5 THz with a minimum of 0.18 dB/cm occurring at 0.481 THz [12]. The loss is mainly limited by the absorption of the host material, but it still suffers from poor isolation [12, 13]. Photonic crystal slab waveguide has been shown to have a loss of about 4 dB/cm between 0.54 and 0.63 THz, but likewise, it suffers from poor isolation and limited bandwidth which is undesirable for multilayer integration and broadband applications [14, 15]. Substrate-integrated waveguide (SIW) has been reported to be broadband and operate to 0.5 THz with loss less than 3.5 dB/cm by selecting different operating modes at different bands, but it is difficult to metalize the vias in SIW - especially when the vias become very small at short wavelengths. Microstructured fibers, including photonic crystal fibers [16-18], antiresonant fibers [19], porous step-index fibers [20], hollow dielectric-lined fibers [21], etc., have all been reported as promising solutions for low-loss quasi-single-mode THz wave-guiding, but they are as yet relatively bulky and hard to integrate with active devices.

In this paper, we show that millimeter and THz waves can be tightly confined in the air core of a hybrid waveguide that is essentially a hollow rectangular waveguide with Bragg reflectors instead of vertical side walls. By avoiding the vertical walls, the waveguide can be fabricated as a photonic crystal structure on a single planar substrate sandwiched between two parallel ground planes, with no metalized via holes. It is, in essence, a flat form of the well-known Bragg fiber [22] that can be realized on a single substrate and so we call it a Substrate Integrated Bragg Waveguide (SIBW). Since it can be realized on a single substrate it is versatile for designing a variety of functional components based on the highly flexible and expandable photonic crystal structure, such as sensors, filters, H-plan horn antennas, lasers and frequency combs, which has already been successfully demonstrated in optical frequencies [23-27]. Based on our theoretical analysis, the SIBW operates in the fundamental $HE_{10}$ mode while suppressing the high-order competing modes using a modal-filtering effect, which allows the waveguide to operate in single-mode fashion over more than one octave. Compared with planar transmission line modes, the vertically polarized $HE_{10}$ mode of the SIBW can be more easily coupled with the free-space Gaussian mode or the $TE_{10}$ mode of a rectangular waveguide, which are the most widely used interfaces for THz-TDS or THz vector network analyzer (VNA) systems, respectively. Besides, the SIBW uses dielectric rather than metal in the vertical wall resulting in a lower propagation loss than that of the classic hollow metallic rectangular waveguide. A proof-of-concept millimeter-wave SIBW was designed and tested in order to verify the transmission properties of the proposed structure. Although it has a relatively low operating frequency, it verifies the physical mechanisms for octave bandwidth single-mode operation and so demonstrates that this is an important candidate for an octave-bandwidth single-mode transmission line for mmW and THz functional components, such as filters, power dividers, and H-plane horn antennas.

* E-mail: gpwang@szu.edu.cn
† E-mail: i.d.robertson@leeds.ac.uk

## II. PRINCIPLE

The schematic diagram of the proposed SIBW is illustrated in Fig. 1(a). The top and bottom layers are parallel metal ground planes, and the middle layer is a substrate that realizes an air-core line-defect 1D photonic crystal Bragg reflector structure. The Bragg reflector consists of periodically alternating layers of high- and low-refractive-index materials. Air is chosen as the low-refractive-index material for its low refractive index and low loss. The other material that forms the Bragg reflector is desired to be high refractive index and low loss, to create a large photonic bandgap and reduce the material absorption loss. High-resistivity silicon is one of the most commonly used materials which has high refractive index and low absorption loss in the THz frequency range [28], and thus it is chosen as the high-refractive-index component of the photonic crystal in our theoretical analysis. The fundamental mode of the proposed SIBW is vertically polarized HE$_{10}$ mode, and its electric field is evenly distributed along the $z$-axis. The height of the SIBW, $h$, does not contribute to the dispersion relations for vertically polarized modes. Therefore, the dispersion relation problem for vertically polarized modes can be simplified into a two-dimensional (2D) problem in the $x$-$y$ plane assuming the third dimension is infinite and uniform. To avoid higher-order competing modes, $h$ should be not greater than $w/2$.

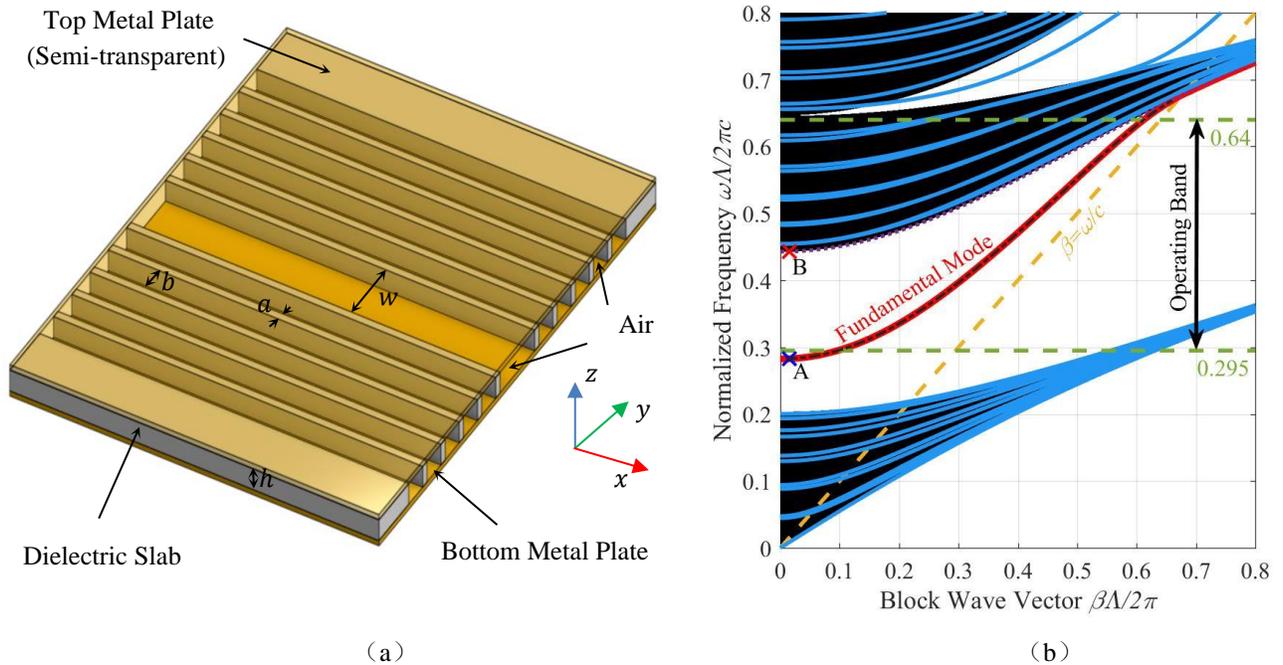

(a)                (b)

FIG 1. (a) Schematic diagram of the proposed SIBW with the key design parameters indicated. $\Lambda = a + b$ is the period of the photonic crystal structure. $w = 1.89\Lambda$ is the width of the line defect in the core region. $h < w/2$ is the thickness of the dielectric slab. (b) Bandgap diagram and dispersion curves for the vertically polarized modes in the SIBW. Here, the dielectric slab is silicon, $a = 0.2276\Lambda$, and $b = 0.7724\Lambda$. The orange dashed line is the light line. The green dashed lines indicate the operating band which is wider than one octave. The black region is the bandpass region for the 1D photonic crystal in which the electromagnetic wave can pass through the periodic structure and leak out. The white region above the light line is the bandgap. The red solid line and the purple dotted line represents the dispersion curves for the fundamental $HE_{10}$ mode and the second-order competing $HE_{20}$ mode, respectively. The blue curves overlaid upon the black region are cladding modes or lossy defect modes whose electric field are mainly distributed in the bulk crystal. The black dot-dashed line overlaid upon the red curve represents the simulated dispersion curve of the fundamental mode using COMSOL.

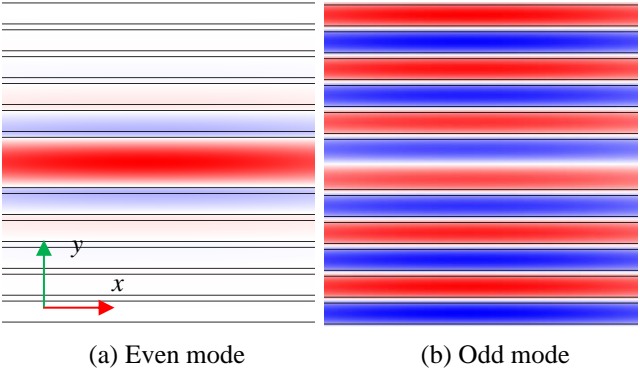

(a) Even mode          (b) Odd mode

FIG. 2. Mode pattern ($E_z$) for points A and B in Fig. 1(b). (a) Point A. $(\beta_n, \omega_n) = (0.015, 0.2828)$; (b) Point B. $(\beta_n, \omega_n) = (0.015, 0.4431)$. Periodic boundaries are applied. Here, $\beta_n = \beta \Lambda/2\pi$ and $\omega_n = \omega \Lambda/2\pi c$.

To maximize the first-order photonic bandgap which supports operating HE$_{10}$ mode, $a$ and $b$ are chosen based on the following quarter-wavelength condition [29]:

$$n_a a = n_b b = \frac{\lambda_c}{4}, \quad (1)$$

where $n_a$ and $n_b$ are the refractive index of the high- and low-refractive-index materials and $\lambda_c$ is the center wavelength of the first-order bandgap when the longitudinal propagation constant $\beta = 0$. As we can see from Eqn. (1), by properly choosing a target operating wavelength ($\lambda_c$) and the materials for the high- and low-refractive-index layers, the geometric parameters of the Bragg reflector structure can change accordingly and the operating band can thereby be scaled to any target operating wavelength. The high-resistivity silicon is dispersionless, and the real part of its complex refractive index is $3.4175 \pm 0.00005$ between 0.5 and 4.5 THz [28]. Thus, in our theoretical design, $n_a = 3.4175$, $n_b = 1$, $a = \frac{n_b}{n_a+n_b}\Lambda = 0.2264\Lambda$ and $b = \frac{n_a}{n_a+n_b}\Lambda = 0.7736\Lambda$. The width of the line defect in the middle of the Bragg reflector waveguide, $w$, greatly affects the dispersion curves of the defect modes, but it barely affects the dispersion curves of the cladding modes. The value of $w$ should be carefully chosen to push the dispersion curve of the competing second-order odd mode to the edge of the photonic bandgap, so the competing mode becomes lossy which stops it from being excited in the SIBW. This creates the well-known modal-filtering effect [30] and allows the SIBW operates over more than one octave.

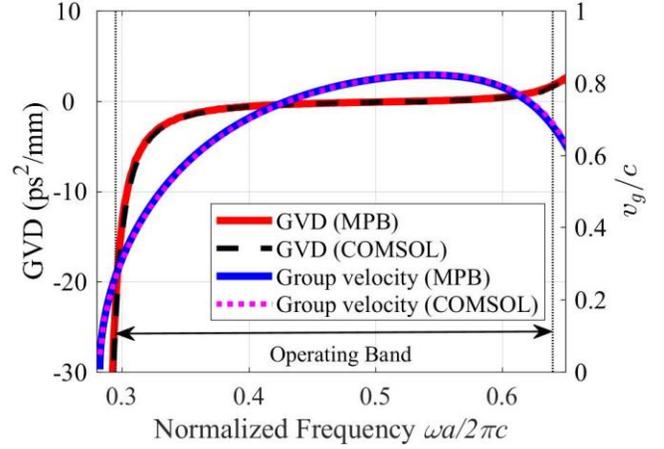

FIG. 3. Group-velocity dispersion ($\partial^2 \beta/\partial \omega^2$) and normalized group velocity ($v_g = \partial \omega/\partial \beta$) of the fundamental mode calculated using MPB and COMSOL.

Dispersion curves for vertically ($z$-axis) linearly polarized modes in the line-defect Bragg reflector waveguide overlaid upon the bandgap diagram are shown in Fig. 1(b). The dispersion curves are numerically calculated using the MIT Photonic-Bands (MPB) package [31], while the photonic bandgap is calculated separately and analytically according to Bloch's theorem [32]. The photonic bandgap obeys the following condition, which provides the constraint which stops the EM wave from propagating through into the periodic claddings from the line defect [32]:

$$|\text{Re}(X_s)| < 1, \quad (2)$$

where

$$X_s = \left[\cos(k_b b) - \frac{i}{2}\left(\frac{\xi_b k_b}{\xi_a k_a} + \frac{\xi_a k_a}{\xi_b k_b}\right)\sin(k_b b)\right]\exp(-ik_a a) \quad (3)$$

Here, $k_i = \sqrt{(n_i k_0)^2 - \beta^2}$ ($i = a, b$) is the lateral propagation constant; $k_0 = \omega/c$ is the vacuum wavenumber; $\beta$ is the longitudinal propagation constant; $n_{eff} = \beta/k_0$ is the effective refractive index of the mode; $\omega$ is the vacuum angular frequency; $c$ is the speed of light in vacuum; and $\xi_i$ is 1 or $1/n_i^2$ ($i = a, b$) corresponding to $S = $ TE or TM modes, respectively.

MPB calculates both the cladding modes, which are mainly distributed in the bulk photonic crystal and the defect modes which are mainly distributed in the central line defect region. As we can see from Fig. 1(b), the dispersion curves for the cladding modes calculated by MPB are coincident with the bandpass region calculated based on Bloch's theorem, validating both methods. The red solid curve corresponds to the desired fundamental HE$_{10}$ mode in SIBW, while the purple dotted line lying at the edge of the photonic bandgap corresponds to the competing and lossy $HE_{20}$ mode. Regarding the $y = 0$ plane, the $HE_{10}$ and $HE_{20}$ modes correspond to the symmetric even mode and antisymmetric odd mode, respectively. The mode pattern of point A for the fundamental $HE_{10}$ mode and point B for the lossy $HE_{20}$ mode in Fig. 1(b) are shown in Fig. 2. As we can see, only the desired fundamental $HE_{10}$ mode can be tightly confined in the line defect region and propagate along the waveguide, while the $HE_{20}$ mode lying at the edge of the photonic bandgap is very lossy and attenuates in the waveguide. This results in the modal-filtering effect [30], which is similar to the method we used in a THz Bragg fiber to suppress unwanted competing modes [33].

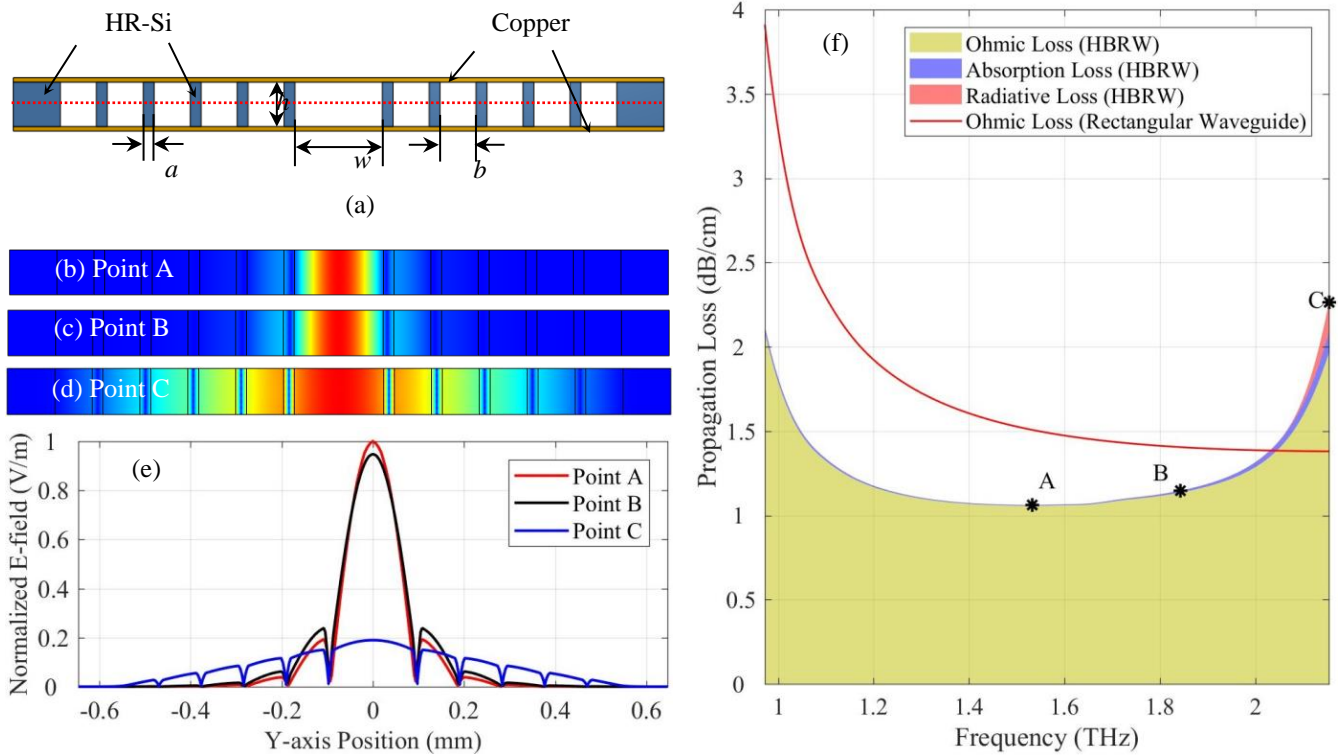

FIG. 4 Structure and characterization of the proposed HCPW. (a) Structure of the SIBW. (b)-(d) Normalized electric field distribution of the desired $HE_{10}$ mode at points A, B, and C shown in Fig. 4(f). The field linearly decreases from red to blue. (e) Normalized electric field along the red dotted line shown in Fig. 4(a) at selected frequency points. (f) Propagation losses of SIBW and hollow metallic rectangular waveguide. The frequencies of points A, B and C are 1.532 THz, 1.843 THz and 2.154 THz, respectively. In the simulation (COMSOL), $a = 0.2264\Lambda \approx 21.7\ \mu m$, $b = 0.7736\Lambda \approx 72\ \mu m$, $w = 1.89\Lambda \approx 176\ \mu m$, and $h = w/2 \approx 88\ \mu m$. The width and height of the hollow metallic rectangular waveguide are 164 μm and 82 μm, respectively.

The frequency range between the two horizontal green dashed lines in Fig. 1(b) gives the operating band of this design. The parts of the dispersion curve for the even mode below the lower dashed line ($\omega_n^l = 0.295$) and above the upper dashed line ($\omega_n^h = 0.64$) are not selected as operating bands since in these frequencies, the guided mode can be relatively lossy and potentially couple to unwanted cladding modes. Here, $\beta_n = \beta\Lambda/2\pi$ and $\omega_n = \omega\Lambda/2\pi c$. The selected operating band gives a ratio bandwidth of $\omega_n^h/\omega_n^l \approx 2.17$, which is wider than one octave. The group-velocity dispersion (GVD) and the normalized group velocity of the fundamental $HE_{10}$ mode are plotted vs. normalized frequency in Fig. 3. Fig. 3 shows that as the frequency increases, the group velocity first increases and then decreases, and the turning point happens to be at $\omega_n = 0.5442$ where the GVD is zero.

## III. SIMULATION

By choosing $\Lambda = 93.1\ \mu m$, the operating band of the above-mentioned SIBW is settled around the frequency range of $(\omega_n^l c/\Lambda, \omega_n^h c/\Lambda) = (0.95\ \text{THz}, 2.06\ \text{THz})$. The eigenvalue problem of SIBW is solved using COMSOL, which is a commercial package based on the finite-element method. The cross-section of the SIBW is used in COMSOL to find the eigenvalues and eigenmodes, as shown in Fig. 4(a). For simplicity, the conductivity of the copper layer is set as $5.8 \times 10^7$ S/m over the frequency of interest and its surface roughness is not considered in the simulation. The dispersion curve of the fundamental $HE_{10}$ mode simulated using COMSOL is plotted in Fig. 1(b) using blue circles, and the GVD and group velocity can be accordingly calculated based on the simulated dispersion relation, as presented using dot-dashed lines in Fig. 3. It can be seen that the COMSOL simulated results agree very well with the MPB calculation which is based on the plane wave expansion method.

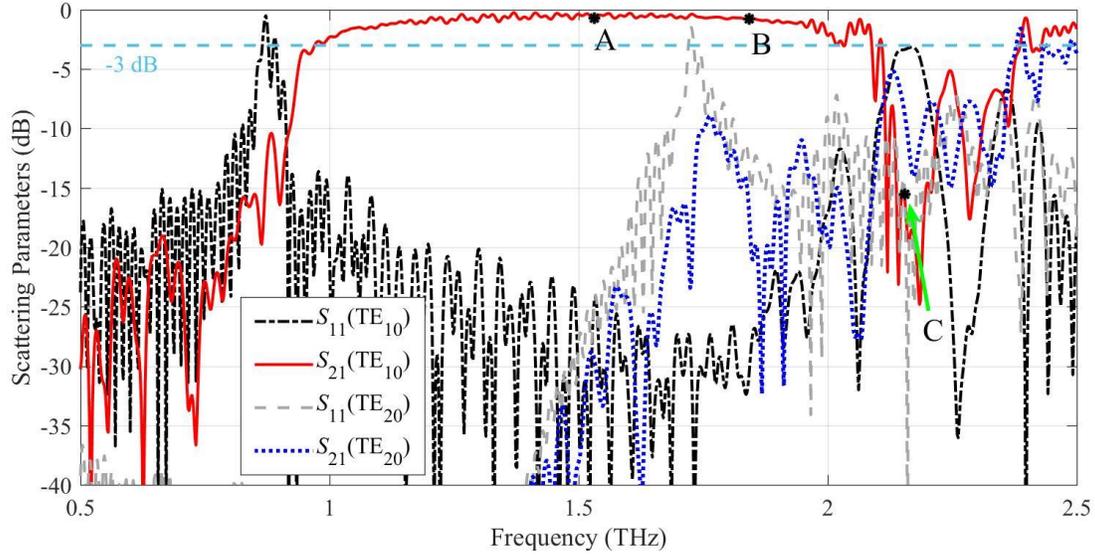

FIG. 5 Scattering Parameters for SIBW when different modes are excited in the hollow metallic rectangular waveguide. The length of the SIBW is 3 mm.

The propagation loss of the SIBW is calculated using the following equation:

$$\alpha \text{ (dB/cm)} = \frac{1}{5\ln 10} \frac{\omega}{c} \text{Im}(n_{\text{eff}}), \qquad (4)$$

where $n_{\text{eff}}$ is the effective mode index of the operating fundamental mode obtained from the COMSOL simulation. Fig. 4(f) presents the simulated propagation loss of the proposed SIBW, as well as the propagation loss of a standard-size WM-164 hollow copper rectangular waveguide for comparison [34]. The propagation loss of the proposed SIBW is below 2 dB/cm over the frequency range wider than one octave from 0.982 to 2.133 THz, which is much lower than the losses of planar transmission lines [7-10], and even lower than that of the standard-size WM-164 hollow copper rectangular waveguide over a broad band wider than the frequency range from 1 to 2 THz. The propagation loss of SIBW consists of Ohmic loss from metal parts, absorption loss from dielectrics, and radiative loss, which are illustrated as the light yellow, purple, and red regions below the total propagation loss line in Fig. 4(f), respectively. It can be seen that the Ohmic loss in the top and bottom copper layers still plays the most significant role. The radiative loss is almost negligible within the operating band, but it increases gradually towards the edges of the operating band, and eventually, the photonic crystal structure loses the ability to confine the electromagnetic wave outside the operating band. The dielectric absorption loss is mainly determined by the absorption coefficient of the dielectric material that constitutes the photonic crystal structure, which is low in this design since the absorption coefficient of high-resistivity silicon is very low over the target frequency range. The dielectric absorption loss is also influenced by the confinement ability of the photonic structure. It can be seen from Fig. 4(f) that, at the high-frequency end of the operating band, the dielectric absorption loss increases with the increase of the radiative loss. This is because when the radiative loss is high, the electromagnetic field penetrates the photonic crystal structure more deeply, and thereby attenuates more significantly by interacting with more dielectric materials. Figs. 4(b) to 4(d) present the field distributions of the desired $HE_{10}$ mode at several representative frequencies, namely the points A, B and C showed in Fig. 4(f). For the points A and B, the field is mainly confined in the air core and the first period of the photonic crystals structure, while for the point C, which is located at the edge of the operating band, the electromagnetic field penetrates further into the photonic crystal structure. Fig. 4(e) presents the normalized electric field along the red dotted line shown in Fig. 4(a) at the representative frequencies, which quantitatively confirms the above results.

Wave propagation in the SIBW was studied using CST Microwave Studio. A hollow metallic rectangular waveguide, whose width (176 μm) and height (88 μm) are equal to those of the defect core of the SIBW, was chosen to feed the SIBW. The single-mode operating frequency range of the hollow metallic rectangular waveguide is between 0.85 THz and 1.70 THz, so at frequencies above 1.70 THz, high order modes, such as $TE_{20}$, $TE_{01}$, etc., can be excited by the hollow metallic rectangular waveguide.

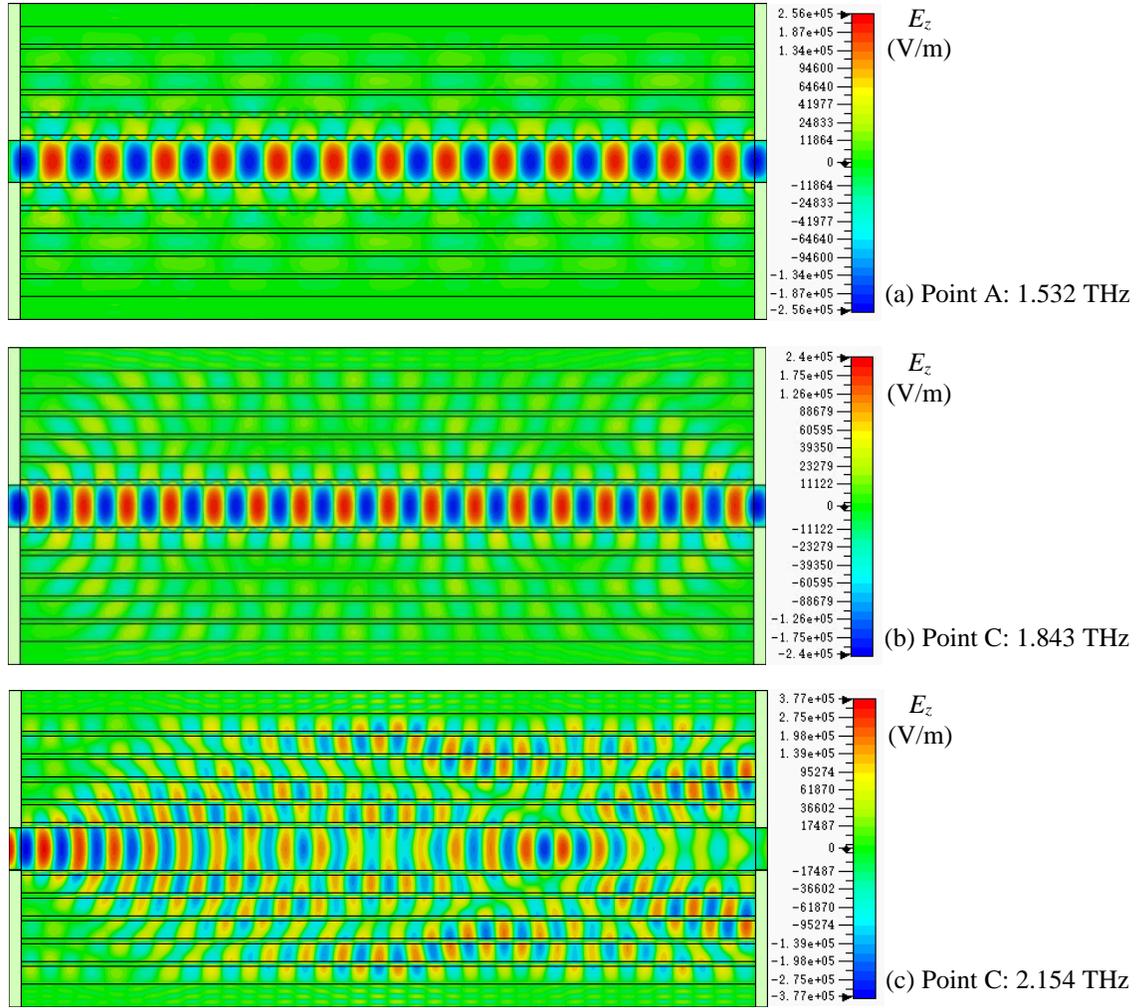

FIG. 6 The z-component of the electric field of SIBW at different representative frequencies which corresponds to the points A, B, and C in Fig. 4, respectively. The color bar is in logarithmic scale.

The scattering parameters for the SIBW when different modes in the hollow metallic rectangular waveguide are excited are shown in Fig. 5. From the $S_{21}$ frequency response, when $TE_{10}$ mode is excited we observe that a transmission window wider than an octave centered at about 1.5 THz is supported in the SIBW. The $S_{11}$ response when the $TE_{10}$ mode is excited indicates that the impedance mismatch between the $TE_{10}$ mode in the hollow metallic rectangular waveguide and the $HE_{10}$ mode in the SIBW is low, since the reflection coefficient is below -14 dB from 1 to 2 THz. When the second-order $TE_{20}$ mode is excited in the hollow metallic rectangular waveguide at around 1.7 THz, the $S_{21}$ rises but it is still below -9 dB owing to the aforementioned modal-filtering effect, and therefore the high-order mode is suppressed. Under the high-order mode ($TE_{20}$) excitation condition, the $S_{11}$ is high at around 1.7 THz which means the impedance mismatch is large at the waveguide transition, and at around 2 THz, both $S_{11}$ and $S_{21}$ are relatively small indicating that the second-order mode is leaky in the SIBW because the electromagnetic field is neither reflected nor transmitted. In addition to the $TE_{20}$ mode, other high-order modes operate at higher frequencies, which is beyond the frequency range of interest from 1 to 2 THz.

The mode patterns of the guided mode in SIBW at several representative frequencies, namely the points A, B, and C shown in Fig. 4, are shown in Fig. 6. It can be seen that the electric field for both frequency point A and point B can be tightly confined in the central defect region, and the operating mode $HE_{10}$ mode is well preserved along the waveguide for both cases. However, for the frequency point C, the electric field penetrates the cladding layers strongly and the guided mode is no longer the single $HE_{10}$ mode. Comparing (a) with (b) in Fig. 6, the electric field at the point B is slightly leakier than that at the point A, and thus the propagation loss at the point B is larger than that at the point A, as shown in Fig. 4. It is common for photonic crystals that the confinement loss is usually smaller at the middle of a photonic bandgap than that at the edge [33].

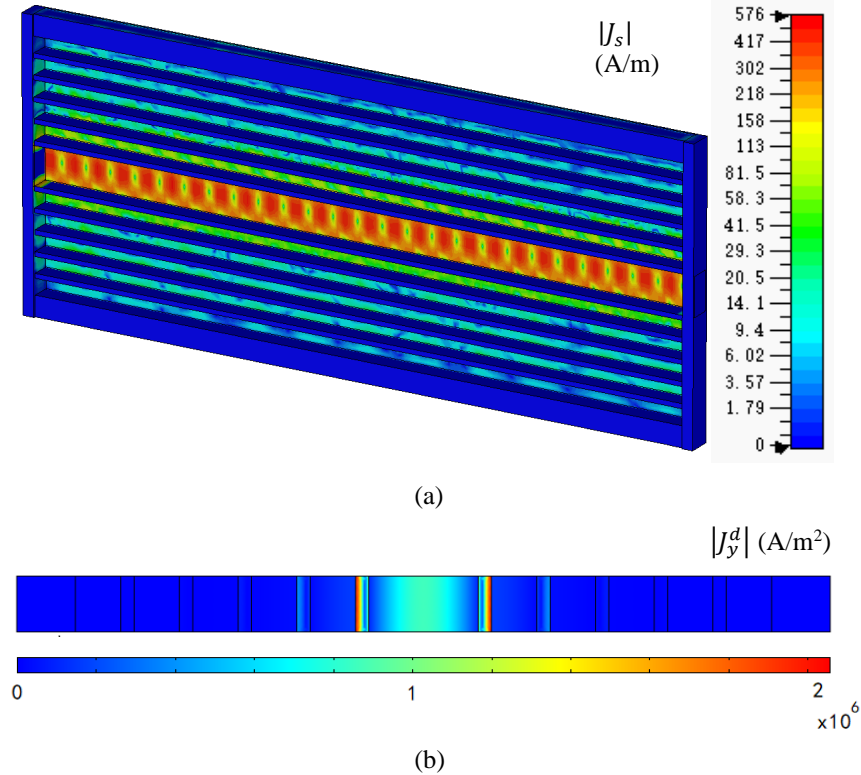

(a)

(b)

FIG. 7 The current distribution of the SIBW. (a) The magnitude of the surface current at 1.532 THz calculated using CST. (b) The magnitude of the y-component of the displacement current density in the cross-section plane of the SIBW calculated using COMSOL.

The distribution of the surface current density in the SIBW at point A is shown in Fig. 7(a). The surface current density is mainly distributed along the central defect region and the first cladding layer, with the remaining small portion being distributed in the outer cladding layers. There is no free current flow between the top and bottom metal plates, in contrast with both the hollow metallic rectangular waveguide and substrate integrated waveguide. Instead, there is displacement current in the dielectric structure that closes the current loop, so generating the magnetic field as shown in Fig. 7(b).

## IV. SCALED EXPERIMENTAL VERIFICATION

To facilitate the fabrication and measurement, a proof-of-concept experimental demonstration of SIBW was performed at millimeter-wave frequencies (ranging from 20 to 45 GHz). Two SIBW samples with different lengths and fed by microstrip and coaxial launchers were fabricated and tested, as shown in Fig. 8. The design principles are the same as those discussed in the principle section. The dielectric material used for the HCPW is Rogers RO3010, which has a dielectric constant of 10.2 and loss tangent of 0.0022 over the frequency range from 8 to 40 GHz. The reason why we use the Rogers RO3010 substrate rather than high-resistivity silicon is that the PCB board is compatible with standard PCB fabrication techniques. The similarities between the RO3010 and the high-resistivity silicon are that they both have relatively high dielectric constant and low absorption loss, which are desirable for constructing the designed photonic crystal structure.

In the two samples shown in Fig. 8, both the microstrip feed lines and the photonic crystal structures were machined using laser direct writing technology with the LPKF ProtoLaser U3. The edges of the laser cutting trajectory were slightly burnt and the burnt dielectric material was left on the structure. The substrates for the microstrip and photonic crystal structure are double-sided copper-clad Rogers RT/Duroid® 5880LZ™ ($\varepsilon_r = 2$) and RO3010™ ($\varepsilon_r = 10.2$), respectively. The thicknesses of the dielectric and copper claddings of both substrates were 0.635 mm and 17.5 μm, respectively. The copper claddings of the RO3010™, which are far thinner than the operating wavelength, were not removed from the dielectric in order to provide extra mechanical support for the delicate photonic crystal structure, and should have negligible impact on the dispersion relation of the SIBW, and only very slightly increase the overall propagation loss, as confirmed by our simulations. The top and bottom copper plates are made of C101 copper. The copper plates were first machined with a DMU 40 CNC milling machine and then manually polished

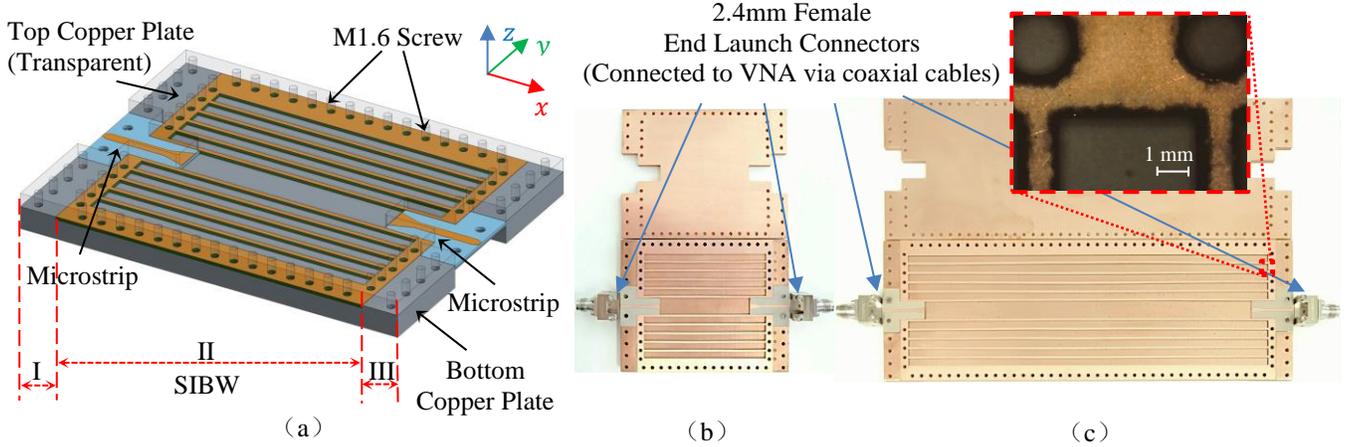

FIG. 8 CAD design and fabricated samples (before assembling) of the SIBW. (a) CAD design. The SIBW in region II is fed by microstrips in regions I and III. (b) Short SIBW with the length between the two microstrip lines $l = 46.54$ mm. (c) Long SIBW with $l = 150.56$ mm. The inserted picture in (c) shows the zoomed view of the photonic crystal structure. In the design, $a = 1$ mm, $b = 3.28$ mm, $w = 8.14$ mm, and $h = 0.671$ mm

by using 500-grade wet-and-dry silicon carbide sandpaper. The bottom plates still had visible residual CNC milling tool marks on the surface, the surface finish of the top plates was smooth.

A two-tier calibration technique was used to characterize the SIBW. The first-tier SOLT calibration was performed using a 2.4mm mechanical calibration kit to place the reference planes to the input interface of the end launch connectors. A second-tier calibration used multiline calibration [35] to de-embed using the complex propagation constant of the SIBW. The two SIBW samples were tested using a vector network analyzer and the scattering parameters are shown in Fig. 9(a). The ripples on the S-parameter responses are mainly due to the impedance mismatches at the interfaces of waveguide transitions and inevitable fabrication imperfections. The reflection coefficients ($S_{11}$) of the two SIBWs are lower than -6 dB (<25%) over the frequency of interest between 20 and 45 GHz, which means that a significant amount of the power was fed into the SIBWs. Besides, the magnitude of the $S_{11}$ parameters for the short and long SIBWs are at the same level, which means that the repeatability of the microstrip to SIBW transitions for both SIBWs are acceptable for multiline calibration. The transmission coefficients ($S_{21}$) for both SIBWs show wideband transmission windows ranging from 20 to 45 GHz which are wider than one octave.

The complex propagation constant of the SIBW can be extracted from the S-parameters shown in Fig. 9(a) using the multiline calibration technique [35]:

$$\gamma = \frac{\ln(\chi)}{\Delta l}, \quad (5)$$

where $\gamma = \alpha + i\beta$ is the complex propagation constant, $\alpha$ is the attenuation factor, $\beta$ is the phase factor, $\Delta l$ is the length difference of the two SIBWs, and $\chi$ is the first eigenvalue of the matrix $T^l(T^s)^{-1}$. Here, $T^s$ and $T^l$ are the $T$-parameters of the short and long SIBWs including the waveguide transition parts, respectively, which can be obtained from the measured S-parameters of the two SIBWs.

The theoretically analyzed and experimentally measured normalized group velocity ($v_g/c = \partial\omega/\partial\beta \,/c$) of the SIBW can be extracted from the dispersion curve calculated using MPB and the measured phase factor based on the Eqn. (5), as shown in Fig. 9(b). Since small ripples on the spectrum of phase factor owing to the small impedance mismatches at discontinuities of the SIBWs or the calibration errors due to the inconsistencies of the two SIBWs have significant impacts on its first-order derivative ($v_g$), reduced frequency points are used when calculating the group velocity so as to analyze the overall trend of the group velocity. The ripples on the curve of the normalized group velocity, which also makes the normalized group velocity larger than one at several frequencies, attributes to the residual and inevitable small impedance mismatches and calibration errors. The 4th-order polynomial fitted measured curve shows the average trends of the normalized group velocity, which agrees with the theoretically analyzed results according to the dispersion relations of SIBW calculated by using MPB. The reason of using 4th-order polynomial curve fitting to the normalized group velocity is due to the factor that the group-velocity should be one order higher than the GVD which follows the trends of a 3rd-order polynomial fitted curve (as shown in Fig. 3). Both the theoretically analyzed and the experimentally measured normalized group velocity rise firstly with increasing frequency and decrease after reaching a maximum. The maximum points for the theoretical and experimental results are 39.73 GHz and 41.05 GHz, respectively, which are close to each other. The turning points in the normalized group velocity respond correspond to the zero GVD point.

To demonstrate the SIBW operates in the desired fundamental $HE_{10}$ mode over the frequency range from 20 GHz to 45 GHz, we measured the GVD and compared it with the theoretical predictions. Fig. 9(c) shows the GVD of the SIBW. The measurement results are calculated using $\partial^2\beta/\partial w^2$ based on the fitted measurement results of group velocity in Fig. 9(b). The theoretically analyzed results are obtained according to the dispersion relations of SIBW calculated by using MPB, which has also been adopted in Fig. 3. It can be seen from Fig. 9(c) that the theoretically analyzed and experimentally measured GVDs are at the same level and follow similar trends. A zero GVD point is observed for both theoretical and experimental results, occurring at 39.73 GHz and 41.05 GHz, respectively.

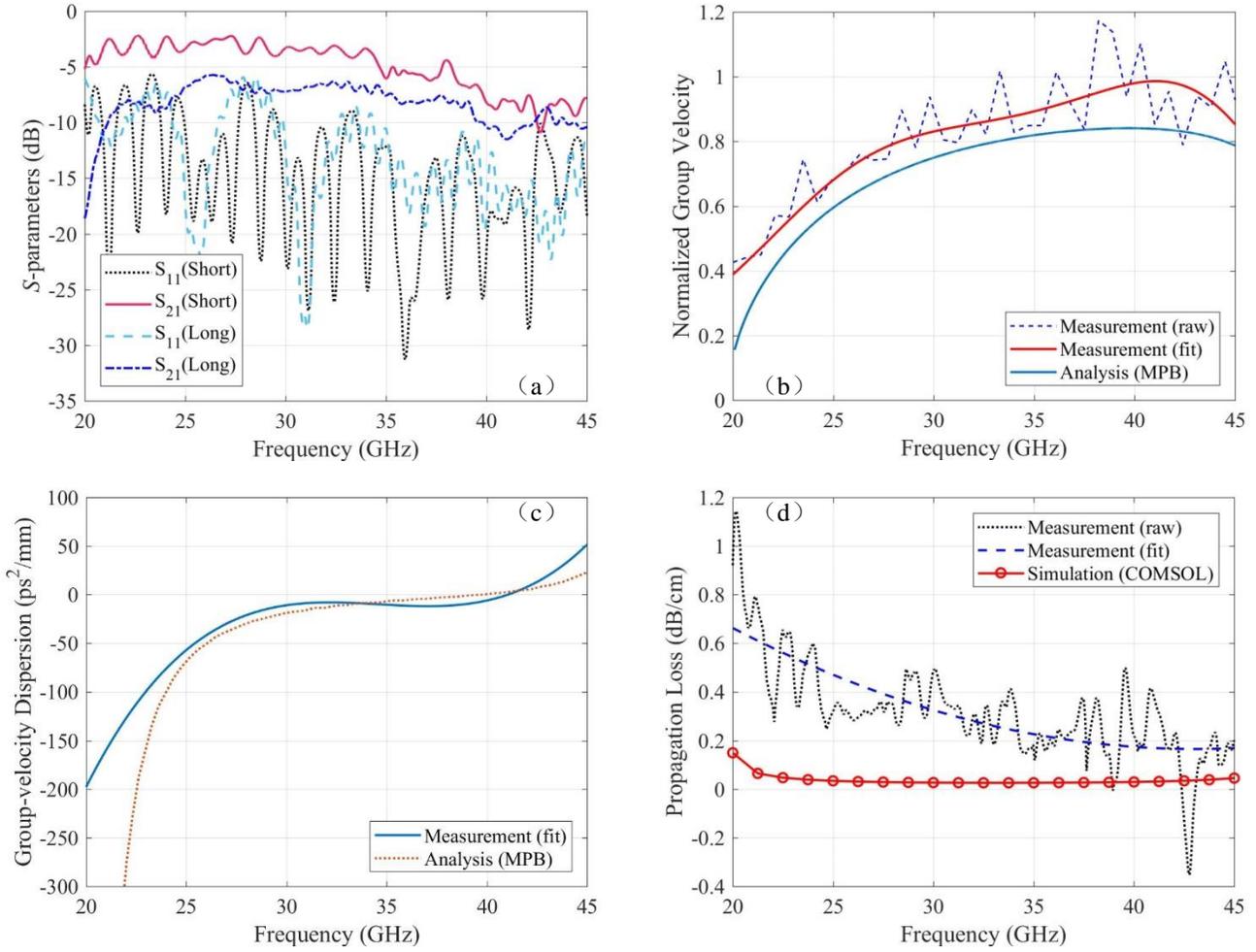

FIG. 9 Characterization of the SIBW. (a) Measured *S*-parameters of the two SIBW samples. One percentage moving window smoothing, a built-in feature of the VNA, is applied to reduce the very high-frequency ripples due to minor impedance mismatches at fabrication imperfections. (b) Normalized group velocity ($v_g/c$). Fourth-order polynomial curve fitting is applied to fit the measured normalized group velocity. (c) Group-velocity dispersion ($\partial^2\beta/\partial w^2$). The measurement result is extracted based on the fitted curve for the normalized group velocity shown in Fig. 9(b). (d) Simulated and measured propagation loss. Second-order polynomial curve fitting is applied to fit the measured propagation loss.

The simulated and measured propagation losses of the SIBW are shown in Fig. 9(d). The measured propagation loss is calculated using $20\log_{10}^e\alpha$ according to the Eqn. (5). The simulated results are obtained based on a COMSOL simulation similar to that used in Fig. 4(f). In general, the measured and fitted simulated results follow similar trends, namely, the propagation losses decrease firstly with increasing frequency and then increases after reaching a minimum. On average, the measured propagation loss of the fabricated SIBW is less than 0.66 dB/cm (compared with 0.15 dB/cm simulated) over the frequency range from 20 to 45 GHz, which is wider than one octave. The measured loss is observed to be greater than the simulated results, which may due to several factors that increase the loss such as surface roughness, oxidation of the copper layer, residual burnt dielectric material, and fabrication or assembly imperfections. At around 42.8 GHz the propagation loss is below zero due to a dip in the response which is a measurement artifact. Comparing with Fig. 9(a), we can see that at the same frequency point, the $S_{21}$ of the short SIBW declines and the $S_{11}$ of the short SIBW rises abnormally, while for the long SIBW, it does not behave in the same way at this frequency. This implies that there is a physical inconsistency between the two SIBWs resulting in multiline calibration errors at around 42.8 GHz, which could be avoided by improving the fabrication and assembly precision.

In summary, we performed a proof-of-concept experimental verification of the guidance mechanism of the proposed SIBW at millimeter-wave frequencies. We found that the measurement results agree with theoretical analysis, which means that the SIBW can operate in single-$HE_{10}$-mode fashion over a bandwidth wider than one-octave, with a GVD passing through zero close to the middle of the operating band. Though the measured propagation loss is slightly higher than the simulated one, as presented in Fig. 9(d), they both follow similar trends. The propagation loss can be further improved and a THz design implemented by using more precise fabrication and assembly techniques, such as deep reactive ion etching (RIE) and wafer-bonding techniques. THz time-domain spectroscopy is a useful tool for wideband characterization of such a waveguide, which will be further discussed in our future work.

## V. CONCLUSIONS

This paper presents a novel substrate integrated Bragg waveguide that can operate in single-$HE_{10}$-mode over more than one octave and in zero group-velocity dispersion fashion around the middle of the operating band. First, we theoretically investigated the SIBW between 1 to 2 THz, and the propagation loss of the proposed SIBW is found to be even lower than that of a classic hollow metallic rectangular waveguide, benefiting from the reduction of Ohmic loss. To facilitate the fabrication and measurement, we performed a proof-of-concept experimental demonstration at millimeter-wave frequencies ranging from 20 to 45 GHz to verify the guidance properties of the proposed SIBW, such as the operating bandwidth, GVD, operating mode, and propagation loss, with consistency between the theoretical analyses (both MPB and COMSOL) and the experimental measurements being observed. Showing many advantages over conventional THz planar waveguides, the proposed SIBW is a promising transmission line for broadband submillimeter-wave and THz systems using functional components or multilayer multichip module technology.

## ACKNOWLEDGMENTS

The author would like to acknowledge Dominic Platt and Lei Shi for their helpful discussion, and also thank Graham Brown and Andrew Pickering from the Engineering and Physical Sciences Research Council (EPSRC) National Facility for Innovative Robotic System for their timely help in fabrication. Besides, the authors also thank financial supports from the National Natural Science Foundation of China (Grant No. 11734012, 51601119, 11574218), the Natural Science Foundation of Shenzhen University (Grant No. 2019006), the EPSRC (Grant No. EP/R00501X/1, EP/P021859/1), and the National Physical Laboratory.

--------------------------------